\documentclass{jnmp_side}

\usepackage{amsmath,graphics,epsfig}

\JNMPnumberwithin{equation}{section}

\newtheorem*{acknowledgments}{Acknowledgments}
\newtheorem{remark}{Remark}

\newtheorem{lemma}{Lemma}

\theoremstyle{definition}

\newtheorem{example}{Example}
\def\openone{\leavevmode\hbox{\small1\kern-3.3pt\normalsize1}}

\def\bbbe{{\Bbb E}}
\def\bbbr{{\Bbb R}}
\def\bbbd{{\Bbb D}}
\def\bbbc{{\Bbb C}}
\def\bbbz{{\Bbb Z}}

\def\conf{\mbox{Conf\,}}

\def\Aut{\mbox{Aut\,}}
\def\re{\mbox{Re\,}}

\def\pb{\boldsymbol{p}}
\def\A{\boldsymbol{A}}
\def\C{\boldsymbol{C}}
\def\D{\boldsymbol{D}}
\def\B{\boldsymbol{B}}
\def\q{\boldsymbol{q}}

\font\dynkfont=cmsy10 scaled\magstep4    \skewchar\dynkfont='60
\def\dynk{\textfont2=\dynkfont}
\def\hr#1,#2;{\dimen0=.4pt\advance\dimen0by-#2pt
              \vrule width#1pt height#2pt depth\dimen0}
\def\vr#1,#2;{\vrule height#1pt depth#2pt}
\def\blb#1#2#3#4#5
            {\hbox{\ifnum#2=0\hskip11.5pt
           \else\ifnum#2=1\hr13,5.4;\hskip-1.5pt
           \else\ifnum#2=2\hr13.5,7.6;\hskip-13.5pt
                        \hr13.5,3.2;\hskip-2pt
                   \else\ifnum#2=3\hr13.7,8.4;\hskip-13.7pt
                                  \hr13,5.4;\hskip-13pt
                                  \hr13.7,2.4;\hskip-2.2pt
                   \else\ifnum#2=7\hr13.7,8.4;\hskip-13.7pt
                                  \hr13.2,6.4;\hskip-13.2pt
                                  \hr13.2,4.4;\hskip-13.2pt
                                  \hr13.7,2.4;\hskip-2.2pt
                   \else\ifnum#2=8\hr13.7,8.4;\hskip-13.7pt
                                  \hr13.2,6.4;\hskip-13.2pt
                                  \hr13.2,4.4;\hskip-13.2pt
                 \hr13.7,2.4;\hskip-8.2pt$\rangle$\hskip-2.0pt
                   \else\ifnum#2=4\hr13.5,7.6;\hskip-13.5pt
                \hr13.5,3.2;\hskip-8pt$\rangle$\hskip-1.8pt
                   \else\ifnum#2=5\hr30,5.4;\hskip-1.5pt
                   \else\ifnum#2=6\hr13.7,8.4;\hskip-13.7pt
                                  \hr13,5.4;\hskip-13pt
               \hr13.7,2.4;\hskip-8.2pt$\rangle$\hskip-2.0pt
                      \fi\fi\fi\fi\fi\fi\fi\fi\fi
                   $#1$
                   \ifnum#4=0
                   \else\ifnum#4=1\hskip-9.2pt\vr22,-9;\hskip8.8pt
                   \else\ifnum#4=2\hskip-10.9pt\vr22,-8.75;\hskip3pt
                                  \vr22,-8.75;\hskip7.1pt
                   \else\ifnum#4=3\hskip-12.6pt\vr22,-8.5;\hskip3pt
                                  \vr22,-9;\hskip3pt
                                  \vr22,-8.5;\hskip5.4pt
                   \else\ifnum#4=5\hskip-9.2pt\vr39,-9;\hskip8.8pt
                                  \fi\fi\fi\fi\fi
                   \ifnum#5=0
                   \else\ifnum#5=1\hskip-9.2pt\vr1,12;\hskip8.8pt
                   \else\ifnum#5=2\hskip-10.9pt\vr1.25,12;\hskip3pt
                                  \vr1.25,12;\hskip7.1pt
                   \else\ifnum#5=3\hskip-12.6pt\vr1.5,12;\hskip3pt
                                  \vr1,12;\hskip3pt
                                  \vr1.5,12;\hskip5.4pt
                   \else\ifnum#5=5\hskip-9.2pt\vr1,29;\hskip8.8pt
                                   \fi\fi\fi\fi\fi

                   \ifnum#3=0\hskip8pt
                   \else\ifnum#3=1\hskip-5pt\hr13,5.4;
                   \else\ifnum#3=2\hskip-5.5pt\hr13.5,7.6;
                                  \hskip-13.5pt\hr13.5,3.2;
                   \else\ifnum#3=3\hskip-5.7pt\hr13.7,8.4;
                                  \hskip-13pt\hr13,5.4;
                                  \hskip-13.7pt\hr13.7,2.4;
                  \else\ifnum#3=7\hskip-5.7pt\hr13.7,8.4;
                                 \hskip-13.2pt\hr13.2,6.4;
                                 \hskip-13.2pt\hr13.2,4.4;
                                  \hskip-13.7pt\hr13.7,2.4;
                  \else\ifnum#3=8\hskip-5.7pt\hr13.7,8.4;
                                 \hskip-13.2pt\hr13.2,6.4;
                                 \hskip-13.2pt\hr13.2,4.4;
                    \hskip-13.9pt$\langle$\hskip-8pt\hr13.7,2.4;
                   \else\ifnum#3=4\hskip-5.5pt\hr13.5,7.6;
                                  \hskip-13.5pt$\langle$\hskip-8.2pt
                                  \hr13.5,3.2;
                  \else\ifnum#3=5\hskip-5pt\hr30,5.4;
                   \else\ifnum#3=6\hskip-5.7pt\hr13.7,8.4;
                                  \hskip-13pt\hr13,5.4;
                       \hskip-13.9pt$\langle$\hskip-8.0pt\hr13.7,2.4;
                                 \fi\fi\fi\fi\fi\fi\fi\fi\fi
                   }}
\def\blob#1#2#3#4#5#6#7{\hbox
{$\displaystyle\mathop{\blb#1#2#3#4#5 }_{#6}\sp{#7}$}}
\def\up#1#2{\dimen1=33pt\multiply\dimen1by#1
                  \hbox{\raise\dimen1\rlap{#2}}}
\def\uph#1#2{\dimen1=17.5pt\multiply\dimen1by#1
                  \hbox{\raise\dimen1\rlap{#2}}}
\def\dn#1#2{\dimen1=33pt\multiply\dimen1by#1
                   \hbox{\lower\dimen1\rlap{#2}}}
\def\dnh#1#2{\dimen1=17.5pt\multiply\dimen1by#1
                    \hbox{\lower\dimen1\rlap{#2}}}

\def\rlbl#1{\kern-8pt\raise3pt\hbox{$\scriptstyle #1$}}
\def\llbl#1{\raise3pt\llap{\hbox{$\scriptstyle #1$\kern-8pt}}}
\def\elbl#1{\kern3pt\lower4.5pt\hbox{$\scriptstyle #1$}}
\def\lelbl#1{\rlap{\hbox{\kern-9pt\raise2.5pt\hbox{{$\scriptstyle #1$}}}}}

\def\wht#1#2#3#4{\blob\circ#1#2#3#4{}{}}

\def\whtd#1#2#3#4#5{\blob\circ#1#2#3#4{#5}{}}

\def\whtu#1#2#3#4#5{\blob\circ#1#2#3#4{}{#5}}
\def\blku#1#2#3#4#5{\blob\bullet#1#2#3#4{}{#5}}
\def\whtr#1#2#3#4#5{\blob\circ#1#2#3#4{}{}\rlbl{#5}}

\def\whtl#1#2#3#4#5{\llbl{#5}\blob\circ#1#2#3#4{}{}}

\def\rwng{\hbox{$\vbox{\offinterlineskip{
  \hbox{\phantom{}\kern6pt{$\circ$}}\kern-2.5pt\hbox{$\Biggr/$}\kern-0.5pt
  \hbox{\phantom{}\kern-5pt$\circ$}\kern-3.0pt\hbox{$\Biggr\backslash$}
  \kern-1.5pt\hbox{\phantom{}\kern6pt{$\circ$}} }}$}}

\def\lwng{\hbox{$\vbox{\offinterlineskip{ \hbox{$\circ$}
  \kern-3.0pt\hbox{\phantom{}\kern6.0pt{$\Biggr\backslash$}}
  \kern-0.5pt\hbox{\phantom{}\kern11pt{$\circ$}}\kern-3.5pt
  \hbox{\phantom{}\kern5.0pt {$\Biggr/$}}\kern-1.0pt\hbox{$\circ$} }}$}}

\def\drwng#1#2#3{\hbox{$\vcenter{ \offinterlineskip{
  \hbox{\phantom{}\kern6pt{$\circ^{\elbl{#3}}$}}
  \kern-2.5pt\hbox{$\Biggr/$}\kern-0.5pt
  \hbox{\phantom{}\kern-5pt$\circ^{ \elbl{#1}}$}
  \kern-3.0pt\hbox{$\Biggr\backslash$}
  \kern-1.5pt\hbox{\phantom{}\kern6pt{$\circ^{\elbl{#2}}$}}  } }$}}

\def\dlwng#1#2#3{\hbox{$\vcenter{\offinterlineskip{ \hbox{$\lelbl{#1}\circ$}
  \kern-3.0pt\hbox{\phantom{}\kern6.0pt{$\Biggr\backslash$}}
  \kern-0.5pt\hbox{\phantom{}\kern11pt{$\lelbl{#2}\circ$}}\kern-3.5pt
  \hbox{\phantom{}\kern5.0pt {$\Biggr/$}}\kern-1.0pt%
   \hbox{$\lelbl{#3}\circ$}}}$}
}

\def\rde#1#2#3{\raisebox{.5pt}{\hbox{\phantom{}\kern-4pt\hbox{$\vcenter
{\offinterlineskip\hbox{
               \raise 4.5pt\hbox{\vrule height0.4pt width13pt depth0pt}
                \kern-1pt\vbox{ \hbox{\drwng{#1}{#2}{#3}}} }}$  }} }}

\def\lde#1#2#3{\raisebox{.5pt}{\hbox{$\vcenter{\offinterlineskip  \hbox{
         \dlwng{#1}{#2}{#3}\kern-5.2pt\lower0.4pt\hbox{$\vcenter{\hrule
 width13pt}$}
               \kern-8pt\phantom{}   }}  $}}}

\def\ldet#1#2#3{\hbox{$\vcenter{\offinterlineskip  \hbox{
               \dlwng{#1}{#2}{#3}\kern14pt\lower0.4pt\hbox{$\vcenter{
          \hskip-20pt\hr13.5,5.7;\hskip-13.5pt\hr13.5,1.3;}$}
               \kern-25.8pt\phantom{}   }}  $}}

\def\rwngb{\hbox{$\vbox{\offinterlineskip{
\hbox{\phantom{}\kern6pt{$\bullet$}}\kern-2.5pt\hbox{$\Biggr/$}\kern-0.5pt
  \hbox{\phantom{}\kern-5pt$\bullet$}\kern-3.0pt\hbox{$\Biggr\backslash$}
  \kern-1.5pt\hbox{\phantom{}\kern6pt{$\bullet$}} }}$}}

\def\lwngb{\hbox{$\vbox{\offinterlineskip{ \hbox{$\bullet$}
  \kern-3.0pt\hbox{\phantom{}\kern6.0pt{$\Biggr\backslash$}}
  \kern-0.5pt\hbox{\phantom{}\kern11pt{$\bullet$}}\kern-3.5pt
  \hbox{\phantom{}\kern5.0pt {$\Biggr/$}}\kern-1.0pt\hbox{$\bullet$} }}$}}

\def\dbrwng#1#2#3{\hbox{$\vcenter{ \offinterlineskip{
  \hbox{\phantom{}\kern6pt{$\bullet^{\elbl{#3}}$}}
  \kern-2.5pt\hbox{$\Biggr/$}\kern-0.5pt
  \hbox{\phantom{}\kern-5pt$\bullet^{ \elbl{#1}}$}
  \kern-3.0pt\hbox{$\Biggr\backslash$}
  \kern-1.5pt\hbox{\phantom{}\kern6pt{$\bullet^{\elbl{#2}}$}}  } }$}}

\def\dblwng#1#2#3{\hbox{$\vcenter{\offinterlineskip{
   \hbox{$\lelbl{#1}\bullet$}
  \kern-3.0pt\hbox{\phantom{}\kern6.0pt{$\Biggr\backslash$}}
  \kern-0.5pt\hbox{\phantom{}\kern11pt{$\lelbl{#2}\bullet$}}\kern-3.5pt
  \hbox{\phantom{}\kern5.0pt
 {$\Biggr/$}}\kern-1.0pt\hbox{$\lelbl{#3}\bullet$}}}$} }

\def\rbde#1#2#3{\hbox{\phantom{}\kern-4pt\hbox{$\vcenter{\offinterlineskip
 \hbox{
               \raise 4.5pt\hbox{\vrule height0.4pt width13pt depth0pt}
                \kern-1pt\vbox{ \hbox{\dbrwng{#1}{#2}{#3}}} }}$  }}  }

\def\lbde#1#2#3{\hbox{$\vcenter{\offinterlineskip  \hbox{
        \dblwng{#1}{#2}{#3}\kern-4.2pt\lower0.4pt\hbox{$\vcenter{\hrule
 width13pt}$}
               \kern-8pt\phantom{}   }}  $}}

\def\eddgiu#1.#2.#3.{\dynk \whtu0100{#1}\whtu1300{#2}\whtu6000{#3}}
\def\eddgid#1.#2.#3.{\dynk \whtd0100{#1}\whtd1300{#2}\whtd6000{#3}}
\def\eddgiid#1.#2.#3.{\dynk  \whtd0300{#1}\whtd6100{#2}\whtd1000{#3}}

\def\eddfiu#1.#2.#3.#4.#5.{\dynk
 \whtu0100{#1}\whtu1100{#2}\whtu1200{#3}\whtu4100{#4}\whtu1000{#5}}

\def\eddfid#1.#2.#3.#4.#5.{\dynk
 \whtd0100{#1}\whtd1100{#2}\whtd1200{#3}\whtd4100{#4}\whtd1000{#5}}

\def\eddfiiu#1.#2.#3.#4.#5.{\dynk
 \whtu0100{#1}\whtu1200{#2}\whtu4100{#3}\whtu1100{#4}\whtu1000{#5}}
\def\eddfiid#1.#2.#3.#4.#5.{\dynk
 \whtu0100{#1}\whtd1200{#2}\whtd4100{#3}\whtu1100{#4}\whtd1000{#5}}

\def\ddanu#1.#2.#3.#4.#5.{\dynk \whtu0100{#1}\whtu1100{#2}\whtu1100{#3}%
                          \cdots\whtu1100{#4}\whtu1000{#5}}
\def\ddanuf#1.#2.#3.#4.{\dynk \whtd0100{#1}\whtu1100{#2}\cdots%
                           \whtu1100{#3}\whtu1000{#4}}
\def\ddanuuf#1.#2.#3.#4.{\dynk \whtu0100{#1}\whtu1100{#2}\cdots%
                           \whtu1100{#3}\whtu1000{#4}}
\def\ddanufd#1.#2.#3.#4.{\dynk \whtd0100{#1}\whtu1100{#2}\cdots%
                           \whtu1100{#3}\whtr1005{#4}}
\def\ddandf#1.#2.#3.#4.{\dynk \whtu0100{#1}\whtd1100{#2}\cdots%
                           \whtd1100{#3}\whtd1000{#4}}
\def\ddanddf#1.#2.#3.#4.{\dynk \whtd0100{#1}\whtd1100{#2}\cdots%
                           \whtd1100{#3}\whtd1000{#4}}
\def\ddandfu#1.#2.#3.#4.{\dynk \whtu0100{#1}\whtd1100{#2}\cdots%
                           \whtd1100{#3}\whtr1050{#4}}
\def\ddcnds#1.#2.#3.#4.#5.#6{\dynk \whtd0200{#1}\whtd4100{#2}%
                          \whtd1100{#3}\cdots%
                           \whtd1100{#4}\whtd1400{#5}\whtd2000{#6}}

\def\eddanu#1.#2.#3.#4.#5.{\dynk \whtu0100{#1}\whtu1100{#2}%
               \up1{\whtr0000{#3}}\cdots\whtu1100{#4}\whtu1000{#5}}
\def\eddand#1.#2.#3.#4.#5.{\dynk \whtd0100{#1}\whtd1100{#2}%
                \up1{\whtr0000{#3}}\cdots\whtd1100{#4}\whtd1000{#5}}
\def\ddand#1.#2.#3.#4.#5.{\dynk \whtd0100{#1}\whtd1100{#2}\whtd1100{#3}%
               \cdots\whtd1100{#4}\whtd1000{#5}}
\def\andfive#1.#2.#3.#4.#5.{\dynk \whtu0100{#1}\whtu1100{#2}\whtu1100{#3}%
                           \whtu1100{#4}\whtu1000{#5}}
\def\andthr#1.#2.#3.{\dynk \whtu0100{#1}\whtu1100{#2}\whtu1000{#3}}

\def\eddanid#1.#2.#3.#4.#5.{\dynk \whtd0200{#1}\whtd4100{#2}%
                           \whtd1100{#3}\cdots\whtd1200{#4}\whtd4000{#5}}
\def\eddanidr#1.#2.#3.#4.#5.{\dynk \whtd0200{#1}\whtd4100{#2}%
                           \cdots\whtd1100{#3}\whtd1200{#4}\whtd4000{#5}}

\def\eddaniid#1.#2.#3.#4.#5.#6.{\hbox{$\vcenter{\hbox
         {\dynk\hbox{$ \lde{#1}{#2}{#3}\whtd1100{#4}\cdots%
          \whtd1400{#5}\whtd2000{#6} $}} }$}}

\def\eddaiii#1.#2.#3.{\dynk\whtd0400{#1}\whtd2200{#2}\whtd4000{#3}}
\def\eddaiiif#1.#2.#3.#4.{\dynk\whtd0400{#1}\whtd2100{#2}%
                    \whtd1200{#3}\whtd4000{#4}}
\def\eddciii#1.#2.#3.{\dynk\whtd0200{#1}\whtd4400{#2}\whtd2000{#3}}

\def\eddbnd#1.#2.#3.#4.#5.#6.{\dynk \lde{#1}{#2}{#3}\whtd1100{#4}\cdots%
                           \whtd1200{#5}\whtd4000{#6}}
\def\eddbndt#1.#2.#3.#4.{\dynk \ldet{#1}{#2}{#3}\hskip-444pt\whtr4000{#4}}

\def\ncddlr#1.#2.#3.#4.#5.{\dynk \whtu0101{#1}\whtu1100{#2}\whtu1100{#3}%
                           \cdots\whtu1100{#4}\whtu1001{#5}}
\def\ncdulr#1.#2.#3.#4.#5.{\dynk \whtd0110{#1}\whtd1100{#2}\whtd1100{#3}%
                           \cdots\whtd1100{#4}\whtd1010{#5}}
\def\ncddrd#1.#2.#3.#4.#5.{\dynk \whtd0100{#1}\whtu1100{#2}\whtu1100{#3}%
                           \cdots\whtu1100{#4}\whtu1005{#5}}
\def\ncddru#1.#2.#3.#4.#5.{\dynk \whtu0100{#1}\whtd1100{#2}\whtd1100{#3}%
                           \cdots\whtd1100{#4}\whtd1050{#5}}
\def\ncddrdu#1.#2.#3.#4.#5.{\dynk \whtu0100{#1}\whtu1100{#2}\whtu1100{#3}%
                         \cdots\whtu1100{#4}\whtu1005{#5}}
\def\ncddrud#1.#2.#3.#4.#5.{\dynk \whtd0100{#1}\whtd1100{#2}\whtd1100{#3}%
                            \cdots\whtd1100{#4}\whtd1050{#5}}
\def\ncddld#1.#2.#3.#4.#5.{\dynk \whtl0105{#1}\whtu1100{#2}\whtu1100{#3}%
                          \cdots\whtu1100{#4}\whtd1000{#5}}
\def\ncddlu#1.#2.#3.#4.#5.{\dynk \whtl0150{#1}\whtd1100{#2}\whtd1100{#3}%
                           \cdots\whtd1100{#4}\whtu1000{#5}}
\def\ncanur#1.#2.#3.#4.#5.{\dynk \whtu0101{#1}\whtu1100{#2}\whtu1100{#3}%
                           \cdots\whtu1100{#4}\whtd1010{#5}}
\def\ncandr#1.#2.#3.#4.#5.{\dynk \whtd0110{#1}\whtd1100{#2}\whtd1100{#3}%
                           \cdots\whtd1100{#4}\whtu1001{#5}}

\def\eddcnd#1.#2.#3.#4.#5.{\dynk \whtd0200{#1}\whtd4100{#2}\whtd1100{#3}
       \cdots \whtd1400{#4}\whtd2000{#5}}

\def\dddnu#1.#2.#3.#4.#5.#6.{\hbox{$\vcenter{\hbox
         {\dynk\hbox{$ \whtu0100{#1}\whtu1100{#2}\cdots%
          \whtu1100{#3}\rde{#4}{#5}{#6} $}}  }$}}
\def\dddnd#1.#2.#3.#4.#5.#6.{\hbox{$\vcenter{\hbox
         {\dynk\hbox{$ \whtd0100{#1}\whtd1100{#2}\cdots%
          \whtd1100{#3}\rde{#4}{#5}{#6} $}} }$}}
\def\dddiv#1.#2.#3.#4.{\hbox{$\vcenter{\hbox
         {\dynk\hbox{$ \whtu0100{#1}\rde{#2}{#3}{#4}
              $}}  }$}}
\def\edddiv#1.#2.#3.#4.{\hbox{$\vcenter{\hbox{\dynk\hbox{$\whtl0100{#1}
\up1{\whtl0001{#2}}\dn1{\whtl0010{#4}}\wht1111\whtr1000{#3} $}}}$}}

\def\edddnu#1.#2.#3.#4.#5.#6.#7.#8.{\hbox{$\vcenter{\hbox
         {\dynk\hbox{$ \lde{#1}{#2}{#3}\whtu1100{#4}\cdots%
          \whtu1100{#5}\rde{#6}{#7}{#8} $}}  }$}}
\def\edddnd#1.#2.#3.#4.#5.#6.#7.#8.{\hbox{$\vcenter{\hbox
         {\dynk\hbox{$ \lde{#1}{#2}{#3}\whtd1100{#4}\cdots%
          \whtd1100{#5}\rde{#6}{#7}{#8} $}} }$}}
\def\edddndf#1.#2.#3.#4.#5.#6.{\hbox{$\vcenter{\hbox
         {\dynk\hbox{$ \lde{#1}{#2}{#3}\rde{#4}{#5}{#6} $}} }$}}

\def\edddnds#1.#2.#3.#4.#5.#6.#7.#8.#9.{\hbox{$\vcenter{\hbox
{\dynk\hbox{$ \lde{#1}{#2}{#3}\whtd1100{#4}\cdot\cdot\whtd1100{#5}\cdot%
      \cdot\whtd1100{#6}\rde{#7}{#8}{#9} $}} }$}}
\def\eddanod#1.#2.#3.#4.#5.#6.{\hbox{$\vcenter{\hbox
         {\dynk\hbox{$ \whtd0200{#1}\whtd4100{#2}\cdots%
          \whtd1100{#3}\rde{#4}{#5}{#6} $}} }$}}

\def\edddniid#1.#2.#3.#4.#5.{\hbox{$\vcenter{\hbox
         {\dynk\hbox{$ \whtd0400{#1}\whtd2100{#2}\whtd1100{#3}\cdots%
          \whtd1200{#4}\whtd4000{#5} $}} }$}}

\def\edddniiu#1.#2.#3.#4.#5.{\hbox{$\vcenter{\hbox
         {\dynk\hbox{$ \blku0200{#1}\whtu2100{#2}\whtu1100{#3}\cdots%
          \whtu1200{#4}\blku2000{#5} $}} }$}}

\def\ddei#1.#2.#3.#4.#5.#6.{\hbox{$\vcenter{\hbox
       {\dynk \whtd0100{#1}\whtd1100{#3}%
       \up1{\whtr0001{#2}}\whtd1110{#4}\whtd1100{#5}\whtd1000{#6}} }$}}

\def\eddei#1.#2.#3.#4.#5.#6.#7.{\hbox{$\vcenter{\hbox
       {\dynk \whtu0100{#1}\whtu1100{#3}%
       \up1{\whtr0011{#2}}\up2{\whtr0001{#7}}\whtd1110{#4}\whtu1100{#5}%
       \whtu1000{#6}} }$}}

\def\ncdddt#1.#2.{\dynk\whtu0400{#1}\whtu2001{#2}}
\def\ncandrt#1.#2.{\dynk\whtd0110{#1}\whtu1001{#2}}
\def\ncanurt#1.#2.{\dynk\whtu0101{#1}\whtd1010{#2}}
\def\ncddet#1.#2.{\dynk\whtu0400{#1}\whtu2000{#2}}
\def\ncddut#1.#2.{\dynk\whtd0400{#1}\whtd2010{#2}}
\def\ncdduot#1.#2.{\dynk\whtd0210{#1}\whtd4000{#2}}
\def\ncddct#1.#2.{\hbox{\dynk\whtu0200{\rotatebox{45}{$\scriptstyle#1$}}%
                    \whtu4000{\rotatebox{45}{$\scriptstyle#2$}}}}
\def\ncddcot#1.#2.{\dynk\whtd0400{\rotatebox{45}{$\scriptstyle#1$}}%
                    \whtd2000{\rotatebox{45}{$\scriptstyle#2$}}}
\def\ncddcst#1.#2.{\dynk\whtu0400{\rotatebox{135}{$\scriptstyle#1$}}%
                    \whtu2000{\rotatebox{135}{$\scriptstyle#2$}}}

\def\rronit#1.{\rotatebox{315}
       {\dynk\whtr5005{\rotatebox{45}{$\scriptstyle #1$}}}}
\def\laronit#1.#2.{\rotatebox{315}{\ncddct#1.#2.}}
\def\raronit#1.#2.{\rotatebox{315}{\ncddcot#1.#2.}}
\def\rarsnit#1.#2.{\rotatebox{225}{\ncddcst#1.#2.}}

\def\ncddd#1.#2.#3.#4.#5.{\dynk\whtu0400{#1}\whtu2100{#2}\whtu1100{#3}%
           \cdots\whtu1100{#4}\whtu1001{#5}}
\def\ncdde#1.#2.#3.#4.#5.{\dynk\whtu0400{#1}\whtu2100{#2}\whtu1100{#3}%
           \cdots\whtu1100{#4}\whtu1000{#5}}
\def\ncdded#1.#2.#3.#4.#5.{\dynk\whtl0400{#1}\whtd2100{#2}\whtd1100{#3}%
           \cdots\whtd1100{#4}\whtd1000{#5}}
\def\ncddeo#1.#2.#3.#4.#5.{\dynk\whtd0100{#1}\whtd1100{#2}\whtd1100{#3}%
           \cdots\whtd1200{#4}\whtd4000{#5}}
\def\ncddeof#1.#2.#3.#4.{\dynk\whtd0100{#1}\whtd1100{#2}%
           \cdots\whtd1200{#3}\whtr4000{#4}}
\def\ncddu#1.#2.#3.#4.#5.{\dynk\whtd0400{#1}\whtd2100{#2}\whtd1100{#3}%
           \cdots\whtd1100{#4}\whtd1010{#5}}
\def\ncdduo#1.#2.#3.#4.#5.{\dynk\whtd0110{#1}\whtd1100{#2}\whtd1100{#3}%
           \cdots\whtd1200{#4}\whtd4000{#5}}
\def\ncddc#1.#2.#3.#4.#5.{\dynk\whtu0200{#1}\whtu4100{#2}\whtu1100{#3}%
           \cdots\whtu1100{#4}\whtu1000{#5}}
\def\ncdddc#1.#2.#3.#4.#5.{\dynk\whtd0200{#1}\whtd4100{#2}\whtd1100{#3}%
           \cdots\whtd1100{#4}\whtd1000{#5}}
\def\ncddcu#1.#2.#3.#4.#5.{\dynk\whtl0200{#1}\whtd4100{#2}\whtd1100{#3}%
           \cdots\whtd1100{#4}\whtr1050{#5}}
\def\ncddcd#1.#2.#3.#4.#5.{\dynk\whtu0200{#1}\whtu4100{#2}\whtu1100{#3}%
           \cdots\whtu1100{#4}\whtu1005{#5}}
\def\ncddco#1.#2.#3.#4.#5.{\dynk\whtd0100{#1}\whtd1100{#2}\cdots%
            \whtd1100{#3}\whtd1400{#4}\whtd2000{#5}}
\def\ncdfr#1.#2.#3.#4.#5.#6.{\ncddc#1.#2.#3.#4.#5.
           \hskip-42.5pt\dynk\rotatebox{315}
          {\whtu5005{\rotatebox{45}{$\scriptstyle #6$}}}}
\def\ncdfrd#1.#2.#3.#4.#5.#6.{\ncdde#1.#2.#3.#4.#5.
           \hskip-42.5pt\dynk\rotatebox{315}
          {\whtr5005{\rotatebox{45}{$\scriptstyle #6$}}}}
\def\ncdfrdc#1.#2.#3.#4.#5.#6.{\ncddc#1.#2.#3.#4.#5.
           \hskip-42.5pt\dynk\rotatebox{315}
          {\whtr5005{\rotatebox{45}{$\scriptstyle #6$}}}}
\def\ncdfrdl#1.#2.#3.#4.#5.#6.{\ncddlu#1.#2.#3.#4.#5.
           \hskip-42.5pt\dynk\rotatebox{315}
          {\whtr5005{\rotatebox{45}{$\scriptstyle #6$}}}}

\def\lronit#1.{\rotatebox{315}
       {\dynk\whtl0550{\rotatebox{45}{$\scriptstyle #1$}}}}
\def\daone#1.#2.{\dynk\whtd0400{#1}\whtd4000{#2}}

\def\ncdfl#1.#2.#3.#4.#5.#6.{\hbox{\lronit#1.\hskip-39.5pt
                   \raisebox{13pt}{$\ddand#2.#3.#4.#5.#6.$}}}
\def\ncdfal#1.#2.#3.#4.#5.{\hbox{\lronit#1.\hskip-39.5pt
                   \raisebox{13pt}{$\ddandf#2.#3.#4.#5.$}}}
\def\ncdfalu#1.#2.#3.#4.#5.{\hbox{\lronit#1.\hskip-39.5pt
                   \raisebox{13pt}{$\ddandfu#2.#3.#4.#5.$}}}

\def\ncdfar#1.#2.#3.#4.#5.{\ddanuf#1.#2.#3.#4.
  \hskip-42.5pt\dynk\rotatebox{315}{\whtr5005{\rotatebox{45}
           {$\scriptstyle #5$}}}}
\def\ncdfaur#1.#2.#3.#4.#5.{\ddanuuf#1.#2.#3.#4.
  \hskip-42.5pt\dynk\rotatebox{315}{\whtu5005{\rotatebox{45}
           {$\scriptstyle #5$}}}}

\def\datwot#1.#2.{\dynk\whtu0700{#1}\whtu8000{#2}}

\def\datwon#1.#2.#3.#4.#5.#6.{\dynk \whtd0200{#1}\whtd4100{#2}%
         \whtd1100{#3}\whtd1100{#4} \cdots\whtd1200{#5}\whtd4000{#6}}
\def\datwono#1.#2.#3.#4.#5.#6.{\dynk \whtd0400{#1}\whtd2100{#2}%
        \whtd1100{#3}\cdots \whtd1100{#4}\whtd1400{#5}\whtd2000{#6}}
\def\datwonl#1.#2.#3.#4.#5.#6.{\dynk \whtd0200{#1}\whtd4100{#2}%
         \whtd1100{#3}\cdots \whtd1100{#4} \whtd1200{#5}\whtd4000{#6}}

\def\ddeii#1.#2.#3.#4.#5.#6.#7.{\hbox{$\vcenter{\hbox
       {\dynk \whtd0100{#1}\whtd1100{#3}%
       \up1{\whtr0001{#2}}\whtd1110{#4}\whtd1100{#5}\whtd1100{#6}%
       \whtd1000{#7}} }$}}

\def\eddeii#1.#2.#3.#4.#5.#6.#7.#8.{\hbox{$\vcenter{\hbox
       {\dynk \whtu0100{#8}\whtu1100{#1}\whtu1100{#3}%
       \up1{\whtr0001{#2}}\whtd1110{#4}\whtu1100{#5}\whtu1100{#6}%
       \whtu1000{#7}} }$}}

\def\ddeiii#1.#2.#3.#4.#5.#6.#7.#8.{\hbox{$\vcenter{\hbox
       {\dynk \whtd0100{#1}\whtd1100{#3}%
       \up1{\whtr0001{#2}}\whtd1110{#4}\whtd1100{#5}\whtd1100{#6}%
       \whtd1100{#7}\whtd1000{#8}} }$}}

\def\eddeiii#1.#2.#3.#4.#5.#6.#7.#8.#9.{\hbox{$\vcenter{\hbox
       {\dynk \whtd0100{#1}\whtd1100{#3}%
       \up1{\whtr0001{#2}}\whtd1110{#4}\whtd1100{#5}\whtd1100{#6}%
       \whtd1100{#7}\whtd1100{#8}\whtd1000{#9}} }$}}

\newcommand\alp{\alpha_}
\newcommand\bet{\beta_}

\begin{document}

%
\renewcommand{\evenhead}{V S Gerdjikov  and G G Grahovski}
\renewcommand{\oddhead}{On Reductions and Real Hamiltonian Forms of Affine Toda
Field Theories}

%
\thispagestyle{empty}

\FirstPageHead{12}{2}{2005}{\pageref{firstpage}--\pageref{lastpage}}{{\bf \tiny{SIDE VI}}}

\copyrightnote{2005}{V S Gerdjikov  and G G Grahovski}

\Name{On Reductions and Real Hamiltonian Forms of Affine Toda
Field Theories}

\label{firstpage}

\Author{Vladimir S GERDJIKOV and Georgi G
GRAHOVSKI}

\Address{Institute for Nuclear Research and Nuclear
Energy, Bulgarian Academy of Sciences, 72 Tsarigradsko chaussee,
1784 Sofia, Bulgaria \\
~~E-mail: gerjikov@inrne.bas.bg \\
~~E-mail: grah@inrne.bas.bg\\[10pt]}

\Date{This article is a part of the special issue titled ``Symmetries and Integrability of Difference Equations (SIDE VI)"}

\begin{abstract}
\noindent  A family of real Hamiltonian forms (RHF) for  the
special class of affine $1+1 $-dimensional Toda field theories is
constructed. Thus the method, proposed in \cite{2} for systems
with finite number of degrees of freedom is generalized to
infinite-dimensional Hamiltonian systems. We show that each of
these RHF is related to a special ${\Bbb Z}_2$-symmetry of the
system of roots for the relevant Kac-Moody algebra. A number of
explicit nontrivial examples of RHF of ATFT are presented.
\end{abstract}

\section{Introduction}

To each simple Lie algebra $\mathfrak{g} $ one can relate a Toda field
theory (TFT) in $1+1 $ dimensions. It allows the Lax representation:
\begin{equation}\label{eq:0.1}
[L,M]=0
\end{equation}
where $L $ and $M $ are first order ordinary differential
operators:
\begin{eqnarray}\label{eq:2.1}
L\psi \equiv \left(  i{d  \over dx } - iq_x(x,t) - \lambda
J_0\right)
\psi (x,t,\lambda )=0, \\
M\psi \equiv \left(  i{d  \over dt } -  {1\over \lambda}
I(x,t)\right) \psi (x,t,\lambda )=0.
\end{eqnarray}
whose potentials take values in $\mathfrak{g} $.
Here $q(x,t) \in \mathfrak{h}$ - the Cartan subalgebra of
$\mathfrak{g}$, $\q(x,t)=(q_1,\dots , q_r) $ is its dual $r
$-component vector, $r=\mbox{rank}\,\mathfrak{ g} $, and
\begin{equation}\label{eq:2.2}
J_0 = \sum_{\alpha \in \pi}^{} E_{\alpha },\qquad I(x,t) =
\sum_{\alpha \in \pi}^{} e^{-(\alpha ,\q(x,t))} E_{-\alpha }.
\end{equation}
By $\pi_{\mathfrak{g}} $ we denote the set of admissible roots of
$\mathfrak{g} $, i.e.  $\pi_{\mathfrak{g}} = \{\alpha _0, \alpha
_1,\dots, \alpha _r\} $ where $\alpha _1,\dots, \alpha _r $ are
the simple roots of $\mathfrak{ g}$ and $\alpha _0 $ is the
minimal root of $\mathfrak{ g} $.  The corresponding TFT is known
as the affine TFT. The Dynkin graph that corresponds to the set of
admissible roots of ${\frak g}$ is called extended Dynkin diagrams
(EDD). The equations of motion are of the form:
\begin{equation}\label{eq:2.3}
{\partial ^2 \q  \over \partial x \partial t } =
\sum_{j=0}^{r} n_j \alpha_j e^{-(\alpha_j ,\q(x,t))},
\end{equation}
where $n_j $ are the minimal positive integers for which
$\sum_{j=0}^{r} n_j\alpha _j=0 $.

The operators $L $ and $M $ are invariant with respect to the
reduction group $\mathcal{ G}_\bbbr\simeq \bbbd_h $ where $h $ is
the Coxeter number of $\mathfrak{ g} $. This reduction group is
generated by two elements satisfying $g_1^h = g_2^2
=(g_1g_2)^2=\openone  $ which allow realizations both as elements
in $\Aut_{\mathfrak{ g}} $ and in $\conf \bbbc $. The invariance
condition has the form  \cite{Mikh}:

\begin{eqnarray}\label{eq:3.1}
C_k(U(x,t,\kappa _k(\lambda ))) = U(x,t,\lambda ), \qquad
C_k(V(x,t,\kappa _k(\lambda ))) = V(x,t,\lambda ),
\end{eqnarray}
where $U(x,t,\lambda ) = -iq_x(x,t) - \lambda J_0$ and
$V(x,t,\lambda ) = -{1\over \lambda} I(x,t) $.  Here $C_k $ are
automorphisms of finite order of $\mathfrak{g} $, i.e.
$C_1^h=C_2^2=(C_1C_2)^2=\openone  $ while $\kappa _k(\lambda ) $
are conformal mappings of the complex $\lambda $-plane. The
algebraic constraints (\ref{eq:3.1}) are automatically compatible
with the evolution. A number of nontrivial reductions of nonlinear
evolution equations can be found in \cite{Como,3}.

\begin{lemma}\label{lem:1}
Let $\mathfrak{g} $ be a simple Lie algebra from one of the
classical series $\A_r $, $\B_r $, $\C_r $ or $\D_r $ and let $h $
be its Coxeter number and $N_0 $ -- the dimension of the typical
representation.  Then the characteristic equation of $J_0$ taken
in the typical representation has the form:
\begin{equation}\label{eq:cp.1}
\zeta ^{r_0} (\zeta ^h-1) =0, \qquad r_0 = N_0-h.
\end{equation}
\end{lemma}

\begin{remark}\label{rem:2}
The constant $r_0 =0$ for $\mathfrak{g}\simeq \A_r, \C_r $; $r_0=1 $ for
$\mathfrak{g}\simeq \B_r$  and $r_0=2 $ for $\mathfrak{g}\simeq \D_r$.
Solving the inverse scattering problem in the last two cases requires
special treatment of the subspaces related to $\zeta =0 $.

\end{remark}

The present paper is an extension of a conference report of one of
us \cite{Elba}. In Section 2 we outline the spectral properties of
the Lax operator.  In Section 3 we generalize the theory of RHF
proposed in \cite{2,Elba} to the ATFT in $1+1$ dimensions. In
Section 4 we show how the involution $\tilde{\mathcal{C}}$ of the
complexified Hamiltonian dynamics (defined in Section 3 below) can
be related to an automorphism $\mathcal{C}^{\#}$ of the
corresponding EDD. This allows one to construct the RHF of ATFT
related to any Kac-Moody algebra. We present here (a
non-exhaustive) list of examples of RHF of ATFT related to several
of the classical series of the Kac-Moody algebras. In deriving
them we make use of the method proposed in \cite{SasKha} for
constructing $\bbbz_p $-symmetries of the EDD; of course our
construction needs only involutions ($\bbbz_2 $-reductions) of
EDD.

\section{The spectral properties of $L $}\label{sec:2}

The reduction conditions (\ref{eq:3.1}) lead to  rather special
properties of the operator $L $. Along with $L $ we will use also
the equivalent system:
\begin{equation}\label{eq:L-t}
Lm(x,t,\lambda ) \equiv i {dm  \over dx } + iq_x m(x,t,\lambda ) -
\lambda [J_0, m(x,t,\lambda )] =0,
\end{equation}
where $m(x,t,\lambda )=\psi (x,t,\lambda )e^{iJ_0x\lambda } $. Combining
the ideas of \cite{GeYa} with the symmetries of the potential
(\ref{eq:3.1}) we can construct a set of $2h $ fundamental analytic
solutions (FAS) $m_\nu (x,t,\lambda ) $ of
(\ref{eq:L-t}) and prove that:

\begin{enumerate}

\item the continuous spectrum $\Sigma  $ of $L $ fills up $2h $ rays
$l_\nu  $ passing through the origin: $\lambda \in l_\nu \colon \arg
\lambda =(\nu -1)\pi /h $;

\item $m_\nu (x,t,\lambda ) $ is a FAS of (\ref{eq:L-t})
analytic with respect to $\lambda  $ in the sector $\Omega _\nu \colon
(\nu -1)\pi /h \leq \arg \lambda \leq \nu \pi /h $ satisfying
$\lim_{\lambda \to\infty } m_\nu (x,t,\lambda )=\openone  $;

\item to each $l_\nu  $ one relates a subalgebra $\mathfrak{g}_\nu
\subset \mathfrak{g} $ such that $\mathfrak{g}_\nu \cap
\mathfrak{g}_\mu =\emptyset $ for $\nu \neq \mu \mod (h) $ and
$\cup_{\nu =1}^{h} \mathfrak{g}_\nu  =\mathfrak{g}$. The symmetry
ensures that each of the subalgebras $\mathfrak{g}_\nu  $ is a
direct sum of $sl(2)$-subalgebras;

\item on $\Sigma  $ the FAS $m_\nu (x,t,\lambda ) $ satisfy
\begin{eqnarray}\label{eq:RHP}
m_\nu (x,t,\lambda ) = m_{\nu -1}(x,t,\lambda ) G_\nu (x,t,\lambda ),
\qquad \lambda \in l_\nu ,\\
\label{eq:G-nu}
G_\nu (x,t,\lambda ) = e^{-i(\lambda J_0x +f(\lambda ))t} G_{0,\nu
}(\lambda ) e^{i(\lambda J_0x +f(\lambda ))t} \in \mathcal{G}_\nu ,
\end{eqnarray}
where $\mathcal{G }_\nu  $ is the subgroup with Lie algebra
$\mathfrak{g }_\nu  $ and  $f(\lambda ) $ is determined by the
dispersion law of the NLEE: $f(\lambda )=\sum_{k=0}^{r}
E_{-\alpha _k}/\lambda  $;

\item the FAS of (\ref{eq:L-t}) satisfy:
\begin{eqnarray}\label{eq:fas}
&& \bar{C}_1 (m_\nu (x,t,\omega \lambda )) = m_{\nu-2
}(x,t,\lambda), \qquad \lambda \in l_{\nu -2},
\end{eqnarray}
where $\bar{C}_1 $ is equivalent to the Coxeter automorphism
\cite{Hum}:
\begin{eqnarray}\label{eq:Cox-di}
\bar{C}_1 (X) \equiv C_1^{-1}XC_1 , \qquad C_1 = e^{{2\pi i
\over h} H_\rho }, \qquad \rho = {1  \over 2 }\sum_{\alpha >0}\alpha ;
\end{eqnarray}
obviously $C_1^h=\openone  $ and $\bar{C}_1(J_0)=\omega^{-1}J_0 $;

\item the FAS $m_\nu (x,t,\lambda ) $ satisfy one of the
following two involutions:
\begin{equation}\label{eq:fas2}
\bar{C}_2 (m_\nu (x,t,\lambda^* ))^\dag = C_2(m_{2h-\nu +2}^{-1}
(x,t,\lambda )),
\end{equation}
where $C_2 $, $C_2^2=\openone  $ is conveniently chosen Weyl group
element, or
\begin{equation}\label{eq:fas3}
(m_\nu (x,t,-\lambda^* ))^* = m_{h-\nu +2} (x,t,\lambda ).
\end{equation}
\end{enumerate}

These relations lead to the following constraints for the sewing
functions $G_{0,\nu }(\lambda ) $ and the minimal set of
scattering data:
\begin{eqnarray}\label{eq:G_nu1}
&& \bar{C}_1 (G_{0,\nu} (\omega\lambda ))= G_{0,\nu -2}(\lambda ),
\\ \label{eq:G_nu2}
&&\bar{C}_2 (G_{0,\nu}^\dag (\lambda^* ))= G_{0,2h-\nu +2}^{-1}
(\lambda ), \\ \label{eq:G_nu3}
&& G_{0,\nu}^* (-\lambda^* )= G_{0,h-\nu +2}(\lambda ).
\end{eqnarray}
If $L $ has no discrete eigenvalues the minimal set of scattering
data is provided by the coefficients of $G_{0,1}(\lambda ),
\lambda \in l_1 $ and $G_{0,2}(\lambda ), \lambda \in l_2 $. All
the other sewing functions $G_{0,\nu }(\lambda ) $ can be
determined from them by applying (\ref{eq:G_nu1}),
(\ref{eq:G_nu2}) or (\ref{eq:G_nu1}), (\ref{eq:G_nu3}).

\section{The real Hamiltonian forms of ATFT}\label{sec:3}

The Lax representations of the ATFT models widely discussed in the
literature (see e.g. \cite{Mikh,OlPerMikh,Olive,SasKha} and the
references therein) are related mostly to the normal real form of the Lie
algebra $\mathfrak{ g} $, see \cite{Helg}. Our aim here is to:

\begin{enumerate}

\item generalize the ATFT to complex-valued fields $\q^\bbbc =
\q^0 + i\q^1$, and

\item describe the family of RHF of these ATFT models.

\end{enumerate}

We also provide a tool to construct new inequivalent RHF's of the ATFT.
This tool is a natural generalization of the one in \cite{2} to
$1+1 $-dimensional systems.  Indeed, the ATFT related to the algebra
$sl(n) $ can be written down as an infinite-dimensional Hamiltonian system
as follows:
\begin{eqnarray}\label{eq:tft.1}
&& {dq_k \over dt } = \{ q_k, H_{\rm ATFT}\}, \qquad {dp_k \over dt } = \{ p_k,
H_{\rm ATFT}\}, \\
\label{eq:tft.2}
&& H_{\rm ATFT} = \int_{-\infty }^{\infty } dx \, \left( {1\over 2 }
(\pb(x,t),\pb(x,t)) + \sum_{k=0}^{r} e^{-(\q(x,t),\alpha_k)}
\right),
\end{eqnarray}
where $\pb = d\q/dt $ and  $\q $ are the canonical momenta
and coordinates satisfying canonical Poisson brackets:
\begin{equation}\label{eq:c-PB}
\{ q_k(x,t) , p_j(y,t)\} = \delta_{jk} \delta (x-y).
\end{equation}
Next we introduce an involution $\mathcal{ C} $ acting on the
phase space $\mathcal{ M} \equiv \{q_k(x), p_k(x)\}_{k=1}^{n} $ as
follows:
\begin{eqnarray}\label{eq:Cc}
&& \mbox{1)} \qquad \mathcal{ C}(F(p_k,q_k)) = F(\mathcal{ C}(p_k),
\mathcal{ C}(q_k)),  \nonumber\\
&& \mbox{2)} \qquad \mathcal{ C}\left( \{ F(p_k,q_k), G(p_k,q_k)\}\right) =
\left\{ \mathcal{ C}(F), \mathcal{ C}(G) \right\} , \nonumber\\
&& \mbox{3)} \qquad \mathcal{ C}(H( p_k,q_k)) = H(p_k,q_k) . \nonumber
\end{eqnarray}
Here $F(p_k,q_k) $, $G(p_k,q_k) $ and the Hamiltonian $H(p_k,q_k) $ are
functionals on $\mathcal{M} $ depending analytically on the fields
$q_k(x,t) $ and $p_k(x,t) $.

The complexification of the ATFT is rather straightforward. The resulting
complex ATFT (CATFT) can be written down as standard Hamiltonian system with
twice as many fields $\q^a(x,t) $, $\pb^a(x,t)  $, $a=0,1 $:
\begin{equation}\label{eq:qp-c}
\pb^\bbbc (x,t) = \pb{}^0(x,t)+i \pb{}^1(x,t), \qquad
\q^\bbbc (x,t)= \q{}^0(x,t)+i \q{}^1(x,t),
\end{equation}
\begin{equation}\label{eq:qp-pb}
\{{q}_{k}^0(x,t), {p}_{j}^0(y,t) \}= - \{{q}_{k}^1(x,t),
{p}_{j}^1(y,t) \} = \delta _{kj} \delta (x-y).
\end{equation}
The densities of the corresponding Hamiltonian and symplectic form
equal
\begin{eqnarray}\label{eq:H_0}
\mathcal{H}_{\rm ATFT}^\bbbc &\equiv & \re \mathcal{H}_{\rm ATFT}
(\pb{}^0+i \pb{}^1, \q{}^0+i \q{}^1) \nonumber\\
&=&  {1\over 2 } (\pb{}^0,\pb{}^0) -{1\over 2 }
(\pb{}^1,\pb{}^1) + \sum_{k=0}^{r} e^{-(\q{}^0,\alpha _k)}
\cos ((\q{}^1,\alpha _k)) ,  \\
\label{eq:ome_0}
\omega^\bbbc &=& (d\pb{}^0\wedge  i d\q{}^0) - (d\pb{}^1\wedge
d \q{}^1).
\end{eqnarray}
The family of RHF then are obtained from the CATFT by imposing an
invariance condition with respect to the involution
$\tilde{\mathcal{ C}} \equiv \mathcal{ C}\circ \ast $ where by
$\ast $ we denote the complex conjugation. The involution
$\tilde{\mathcal{ C}} $ splits the phase space $\mathcal{ M}^\bbbc
$ into a direct sum $\mathcal{ M}^\bbbc \equiv {\cal M}_+^\bbbc
\oplus \mathcal{M}_-^\bbbc$ where
\begin{equation}\label{eq:M-c}
\mathcal{M}_+^\bbbc = \mathcal{ M}_0 \oplus i \mathcal{ M}_1, \qquad
\mathcal{M}_-^\bbbc = i\mathcal{ M}_0 \oplus  \mathcal{ M}_1,
\end{equation}
The phase space of the RHF is $\mathcal{ M}_\bbbr \equiv
\mathcal{M}_+^\bbbc $.  By $\mathcal{ M}_0 $ and $\mathcal{ M}_1 $ we
denote the eigensubspaces of $\mathcal{ C} $, i.e.
$\mathcal{C}(u_a)=(-1)^a u_a $ for any $u_a\in {\cal M}_a $.

Thus to each involution $\mathcal{ C} $ satisfying 1) - 3) one
can relate a RHF of the ATFT.  Due to the condition 3) $\mathcal{C} $
must preserve the system of admissible roots of $\mathfrak{g} $; such
involutions can be constructed from the $\bbbz_2 $-symmetries of the
extended Dynkin diagrams of $\mathfrak{ g} $ studied in \cite{SasKha}.

In fact one can relate ATFT models not only to the algebra $sl(n) $
\cite{Mikh}, but also to each Kac-Moody algebra
\cite{OlPerMikh,Olive,1}; for the theory of Kac-Moody algebras see
\cite{Kac}.  The construction is as follows. Let
$\pi_{\mathfrak{g}}=\{\alpha _0, \alpha _1,\dots,\alpha _r\} $  be the set
of admissible roots of $\mathfrak{g} $. Then the
corresponding ATFT model is given by:
\begin{eqnarray}\label{eq:H-g}
H_{\mathfrak{g}} &=& \int_{-\infty }^{\infty }dx\,
\mathcal{H}_{\mathfrak{g}}(x,t), \qquad
\mathcal{H}_{\mathfrak{g}}(x,t)= { 1\over 2} (\pb(x,t),
\pb(x,t))+\sum_{k=0}^{r} n_ke^{-(\q(x,t),\alpha _k)}  ,\\
\label{eq:ome-g}
\Omega _{\mathfrak{g}} &=& \int_{-\infty }^{\infty } dx\,
\omega_{\mathfrak{g}} (x,t), \qquad
\omega_{\mathfrak{g}} (x,t)=( \delta \pb(x,t)\wedge \delta \q(x,t)),
\end{eqnarray}
In what follows we will need also the integer coefficients $n_k $ that
provide the decomposition of the $\alpha _0 $ over the simple roots of
$\mathfrak{g} $:
\begin{equation}\label{eq:n_k}
-\alpha _0 = \sum_{k=1}^{r} n_k\alpha _k.
\end{equation}
Just like for the $sl(n) $ case one can consider the
complexification:
\begin{eqnarray}\label{eq:HC-g}
H_{\mathfrak{g}}^\bbbc &=& \int_{-\infty }^{\infty }dx\,
\mathcal{H}_{\mathfrak{g}}^\bbbc (x,t), \qquad
\mathcal{H}_{\mathfrak{g}}^{\bbbc}(x,t) =
{ 1 \over 2} (\pb^{0}(x,t),\pb^{0}(x,t)) - { 1 \over 2}
(\pb^{1}(x,t),\pb^{1}(x,t)) \nonumber\\
&+& \sum_{k=0}^{r}n_k e^{-(\q^{0}(x,t),\alpha _k)}
\cos (\q^{1}(x,t),\alpha _k)  ,\\
\label{eq:omeC-g}
\Omega_{\mathfrak{g}}^\bbbc &=& \int_{-\infty }^{\infty } dx\,
\omega_{\mathfrak{g}}^\bbbc (x,t), \qquad \omega _{\mathfrak{g}}^{\bbbc} =
(\delta \pb^{0}(x,t)\wedge \delta \q^{0}(x,t)- (\delta
\pb^{1}(x,t)\wedge \delta \q^{1}(x,t) .
\end{eqnarray}
In order to construct the RHF of the ATFT we will make use of
specifically constructed involutions $\mathcal{C} $ of the phase
space $\mathcal{M}_{\mathfrak{g}} \equiv \{ \pb(x,t),\q(x,t)\}$.
An important fact in our construction will be that to each
involution $\mathcal{C} $ there exist a dual involution
(involutive automorphism) $\mathcal{C}^\# $ of $\mathfrak{g} $.
Property 3) above along with the definition (\ref{eq:H-g}) of
$\mathcal{H}_{\mathfrak{g}} $ implies that $\mathcal{C}^\# $
preserves the set of admissible roots $\pi_{\mathfrak{g}} $. For
brevity below we will skip the dependence of $\pb $ and $\q $ on
$x $ and $t $.

Indeed, the condition 3) above requires that:
\begin{equation}\label{eq:*1}
(\mathcal{C}(\q ),\alpha ) = (\q , \mathcal{C}^\# (\alpha )), \qquad
\alpha \in \pi_{\mathfrak{g}},
\end{equation}
and therefore we must have $\mathcal{C}(\pi_{\mathfrak{g}}) =
\pi_{\mathfrak{g}} $.

The relation (\ref{eq:*1}) defines uniquely the relation between
$\mathcal{C} $ and $\mathcal{C}^\# $. Using $\mathcal{C}^\#  $ we can
split the root space $\bbbe^n $ into direct sum $\bbbe^n =\bbbe_+\oplus
\bbbe_- $ of two eigensubspaces of $\mathcal{C}^\# $. Taking the average
of the roots $\alpha _j $ with respect to  $\mathcal{C}^\# $ we get:
\begin{equation}\label{eq:*2}
\beta _j = {1  \over 2 } (\alpha _j+\mathcal{C}^\#(\alpha _j)), \qquad
j=0,\dots, n_+=\dim \bbbe_+.
\end{equation}
By construction the set $\{\beta _0,\beta_1,\dots , \beta _{n_+}
\}$ will be a set of admissible roots for some Kac-Moody algebra
with rank $n_+ $. Graphically each set of admissible roots can be
represented by an extended Dynkin diagrams. Therefore one can
relate an automorphism $\mathcal{C}^\# $ to each $\bbbz_2 $
symmetry of the extended Dynkin diagram.

The splitting of $\bbbe^n $ naturally leads to the splittings of the
fields:
\begin{eqnarray}\label{eq:*3}
\pb =\pb^+  +\pb^- , \qquad \q =\q^+  +\q^- ,
\end{eqnarray}
where $\pb^+ ,\q^+  \in \bbbe_+$ and $\pb^- ,\q^-  \in \bbbe_-$. If
we also introduce:
\begin{equation}\label{eq:*4}
\gamma _j = {1  \over  2} (\alpha _j - \mathcal{C}^\#(\alpha _j)), \qquad
j=0,\dots , n_- =\dim \bbbe_-.
\end{equation}
\begin{remark}\label{rem:A}
In enumerating the vectors $\beta _j $ and $\gamma _j $ we take into
account only those values of $j $ which lead to non-vanishing results in
the r.h.sides of (\ref{eq:*2}) and (\ref{eq:*4}).
\end{remark}
Then applying the ideas of \cite{2} we obtain the following result
for the RHF of the ATFT related to $\mathfrak{g} $:
\begin{eqnarray}\label{eq:*H-g}
\mathcal{H}^{\bbbr}_{\mathfrak{g}} &=&
{1 \over 2} (\pb^+ , \pb^+ ) -{1 \over 2} (\pb^- , \pb^- ) +
\sum_{k=0}^{r} n_k' e^{-(\q^+ ,\beta _k)} \cos (\q^- ,\gamma _j),\\
\label{eq:*ome-g}
\omega ^{\bbbr}_{\mathfrak{g}} &=&
(\delta \pb^+  \wedge \delta \q^+) -(\delta \pb^-  \wedge \delta \q^-) ,
\end{eqnarray}
where $n_k' $ are the minimal positive integers for which
\begin{equation}\label{eq:n_k'}
-\beta _0 = \sum_{k=1}^{r} n_k'\beta _k.
\end{equation}
We can consider particular cases of the  models (\ref{eq:*H-g}) by
imposing on them the constraints $\pb^- =0 $ and $\q^- =0 $. This
leads to an ATFT related to a Kac-Moody algebra $\mathfrak{g}' $
of rank $r' $ whose system of admissible roots $\pi' =\{\beta
_0,\beta _1,\dots , \beta _{r'}\} $ is obtained from $\pi $ by
averaging with respect to the involution $\mathcal{C} $. For each
of the examples listed in the next Section we provide both sets of
admissible roots.

The RHF of ATFT are more general integrable systems than the models
described in \cite{1,SasKha} which involve only the fields $\q^+ $, $\pb^+
$ invariant with respect to $\mathcal{C} $.

\section{Examples}

In this Section we provide several examples of RHF of ATFT related to
Kac-Moody algebras from the series $\A_n^{(a)} $ and $\D_{n}^{(a)} $ where
the height $a $ can be either 1 or 2. We show that the involutions
$\mathcal{C} $ in all these cases are dual to $\bbbz_2 $-symmetries  of
the extended Dynkin diagrams derived in \cite{SasKha}.

The dynamical variables for the corresponding ATFT systems related to the
type $\A_r $ Kac-Moody algebras are provided as $r+1 $-component
vector-valued functions $\pb(x,t) $ and $\q(x,t) $ restricted by the
condition
\begin{equation}\label{eq:rest}
\sum_{k=1}^{r+1} p_{k}(x,t) =\sum_{k=1}^{r+1} q_{k}(x,t) =0.
\end{equation}
This is related to the fact  that the root systems of type $\A_r $
Kac-Moody algebras are conveniently embedded as $r $-dimensional subspace
of $\bbbe^{r+1} $ orthogonal to the vector $e_1+e_2+\dots + e_{r+1} $.

Below we list several examples which illustrate the procedure outlined
above and display new types of ATFT.

\begin{example}\label{exa:1}

We choose $\mathfrak{g}\simeq \A_{2r-1}^{(1)} $ and fix up the involution
$\mathcal{C} $ by:
\begin{eqnarray}\label{eq:C_1}
&&\mathcal{C} (q_k) =-q_{2r+1-k}, \qquad \mathcal{C} (p_k)
=-p_{2r+1-k},\qquad k=1,\dots , r,
\end{eqnarray}
The coordinates in $\mathcal{M}_\pm $ are given by:
\begin{eqnarray}\label{eq:p-pm1}
q_{k}^{\pm} = {1 \over \sqrt{2}} (q_k\mp q_{2r+1-k}),\qquad
p_{k}^{\pm} = {1 \over \sqrt{2}} (p_k\mp p_{2r+1-k}),
\end{eqnarray}
where $k=1, \dots , r $, i.e., $\dim \mathcal{M}_+= 2r$, $\dim
\mathcal{M}_-=2r-2$.

This involution induces an involution $\mathcal{C}^\# $ of the Kac-Moody
algebra $\A_{2r-1}^{(1)} $ which acts on the root space as follows,
 (see fig.~\ref{fig:1}):
\begin{eqnarray}\label{eq:E1.2}
\mathcal{C}^\# e_k =-e_{2r+1-k}, \qquad k=1,\dots, 2r; \\
\mathcal{C}^\#\alpha _k =\alpha _{2r-k}, \qquad k=1,\dots r-1; \qquad
\mathcal{C}^\# \alpha _0 =\alpha _0, \qquad \mathcal{C}^\#\alpha _r =
\alpha _r.
\end{eqnarray}
The involution $\mathcal{C}^\# $ splits the root space
$\bbbe^{2r-1} $ into a direct sum of its eigensubspaces:
$\bbbe^{2r-1}=\bbbe_+\oplus \bbbe_- $ with $\dim \bbbe_+=r $ and
$\dim \bbbe_-=r-1 $. The restriction of $\pi $ onto $\bbbe_+ $
leads to the admissible root system $\pi'=\{ \beta _0,\dots, \beta
_r\} $ of $\C_{r}^{(1)} $:
\begin{equation}\label{eq:bet-1}
\beta _k={1 \over 2} (\alpha _k+\mathcal{C}^\#\alpha _k ), \qquad
k=1,\dots, r-1; \qquad \beta _0=\alpha _0, \qquad \beta _r=\alpha _r.
\end{equation}
The subspace $\bbbe_- $ is spanned by
\begin{equation}\label{eq:gam-1}
\gamma _k={1 \over 2} (\alpha _k-\mathcal{C}^\#\alpha _k ), \qquad
k=1,\dots, r-1.
\end{equation}
Then the densities $\mathcal{H}_{1}^\bbbr$, $\omega _{1}^\bbbr$
for the RHF of AFTF equal:
\begin{eqnarray}\label{eq:H1}
\mathcal{H}_{1}^\bbbr  &=& {1\over 2} \sum_{k=1}^{r-1}
(p_{k}^{+}{}^2 -  p_{k}^{-}{}^2)+\sum_{k=1}^{r-1} 2 e^{ (q_{k+1}^+
-q_k^+)/\sqrt{2}} \cos \left( {q_{k+1}^- -q_k^- \over \sqrt{2}}\right)\\
&+& e^{\sqrt{2}q_1^+} + e^{-\sqrt{2}q_{r}^+}   ,\nonumber \\
\label{eq:ome1}
\omega _{1}^\bbbr &=& \sum_{k=1}^{r} \delta p_k^+(x)\wedge \delta q_k^+(x)
- \sum_{k=1}^{r+1} dp_k^-\wedge dq_k^- .
\end{eqnarray}
where $\pb_k^\pm = d\q_k^\pm /dt $. If we put $\q^- = 0 $ and $\pb^-=0 $
we get the reduced ATFT related to the Kac-Moody algebra $\C_{r+1}^{(1)} $
\cite{SasKha}.

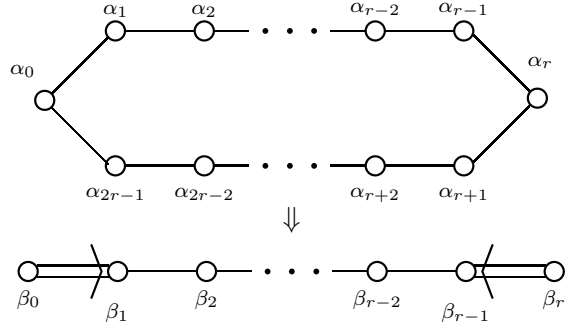
\begin{figure}
\begin{picture}(300,140)
\put(111,83){$\lronit{\alp{0}}.$}
\put(142,96){$\ncdfaur{\alp{1}}.{\alp{2}}.{\alp{r-2}}.
     {\alp{r-1}}.{\alp{r}}.$}
\put(142,45){$\ddanddf{\alp{2r-1}}.{\alp{2r-2}}.
     {\alp{r+2}}.{\alp{r+1}}.$}
\put(222,27){$\Downarrow$}
\put(109,5){$\ddcnds{\bet{0}}.{\bet{1}}.{\bet{2}}.{\bet{r-2}}.
     {\bet{r-1}}.{\bet{r}}.$}
\end{picture}
\caption{$\A_{2r-1}^{(1)}\rightarrow \C_r^{(1)}$}
\label{fig:1}
\end{figure}

\end{example}

\begin{example}\label{exa:2}

Choose $\mathfrak{g}\simeq \A_{2r+1}^{(1)}, r>2 $ and fix up the
involution $\mathcal{C} $ by:
\begin{eqnarray}\label{eq:C_2}
&&\mathcal{C} (q_k) =-q_{2r+2-k}, \qquad \mathcal{C} (p_k)
=-p_{2r+2-k},\qquad k=1,\dots , r, \nonumber\\
&& \mathcal{C} (q_{r+1})=-q_{r+1}, \qquad \mathcal{C} (p_{r+1})
=-p_{r+1}.
\end{eqnarray}
Here the coordinates in $\mathcal{M}_\pm $ are given by:
\begin{eqnarray}\label{eq:p-pm2}
q_{k}^{\pm} = {1 \over \sqrt{2}} (q_k\mp q_{2r+2-k}),\qquad
p_{k}^{\pm} = {1 \over \sqrt{2}} (p_k\mp p_{2r+2-k}), \qquad
k=1, \dots , r , \\
q_{r+1}^- = q_{r+1}, \qquad  p_{r+1}^- = p_{r+1},\nonumber
\end{eqnarray}
i.e., $\dim \mathcal{M}_+= 2r$ and $\dim\mathcal{M}_-=2r+2 $.

This involution induces an involution $\mathcal{C}^\# $ of the Kac-Moody
algebra $\A_{2r+1}^{(1)} $ which acts on the root space as follows
(see fig.~\ref{fig:2}):
\begin{eqnarray}\label{eq:E2.2}
\mathcal{C}^\# e_k =-e_{2r+2-k}, \qquad k=1,\dots, r; \quad k=r+2,\dots,
2r+1; \nonumber\\
\mathcal{C}^\# e_{2r+2} =-e_{2r+2}, \qquad \mathcal{C}^\# e_{r+1}
=-e_{r+1}, \\
\mathcal{C}^\#\alpha _k =\alpha _{2r+1-k}, \qquad k=0,\dots 2r+1;
\end{eqnarray}
The involution $\mathcal{C}^\# $ splits the root space $\bbbe^{2r}
$ into a direct sum of its eigensubspaces:
$\bbbe^{2r+1}=\bbbe_+\oplus \bbbe_- $ with $\dim \bbbe_+=r $ and
$\dim \bbbe_-=r+1 $. The restriction of $\pi $ onto $\bbbe_+ $
leads to the admissible root system $\pi'=\{ \beta _0,\dots, \beta
_r\} $ of $\D_{r+1}^{(2)} $:
\begin{equation}\label{eq:bet-2}
\beta _k={1 \over 2} (\alpha _k+\mathcal{C}^\#\alpha _k ), \qquad
k=0,\dots, r;
\end{equation}
The subspace $\bbbe_- $ is spanned by
\begin{equation}\label{eq:gam-2}
\gamma _k={1 \over 2} (\alpha _k-\mathcal{C}^\#\alpha _k ), \qquad
k=0,\dots, r.
\end{equation}

Then the densities $\mathcal{H}_{2}^\bbbr$, $\omega
_{2}^\bbbr$  for the RHF of AFTF equal:
\begin{eqnarray}\label{eq:H2}
&& \mathcal{H}_{2}^\bbbr  = {1\over 2} \sum_{k=1}^{r}
p_{k}^{+}{}^2 - {1\over 2} \sum_{k=1}^{r+1} p_{k}^{-}{}^2 + 2
e^{-q_{r-1}^+/\sqrt{2}} \cos \left(
{q_{r-1}^- \over \sqrt{2} } - q_{r+1}^-\right) \nonumber\\
&& +\sum_{k=1}^{r-1} 2 e^{ (q_{k+1}^+ -q_k^+)/\sqrt{2}} \cos
\left( {q_{k+1}^- -q_k^- \over \sqrt{2}}\right)+ 2
e^{q_1^+/\sqrt{2}} \cos \left( {q_{1}^- \over \sqrt{2} } -
q_{r+1}^-\right) .\nonumber \\
\label{eq:ome2}
&& \omega _{2}^\bbbr = \sum_{k=1}^{r} \delta p_k^+\wedge \delta q_k^+ -
\sum_{k=1}^{r+1} \delta p_k^-\wedge \delta q_k^-
\end{eqnarray}
where $\pb_k^\pm = d\q_k^\pm /dt $. If we put $q_j^- = 0
$ and $p_j^-=0 $ we get the reduced ATFT related to the Kac-Moody
algebra $\D_{r+1}^{(2)} $ \cite{SasKha}.

\begin{figure}
\begin{picture}(300,100)
\put(120,80){$\ncddlr{\alp{0}}.{\alp{1}}.{\alp{2}}.
         {\alp{r-1}}.{\alp{r}}.$}
\put(120,46){$\ncdulr{\alp{2r+1}}.{\alp{2r}}.{\alp{2r-1}}.
     {\alp{r+2}}.{\alp{r+1}}.$}
\put(218,27){$\Downarrow$}
\put(120,5){$\edddniid{\bet{0}}.{\bet{1}}.{\bet{2}}.
         {\bet{r-1}}.{\bet{r}}.$}
\end{picture}
\caption{$\A_{2r+1}^{(1)}\rightarrow \D_{r+1}^{(2)}$}
\label{fig:2}
\end{figure}
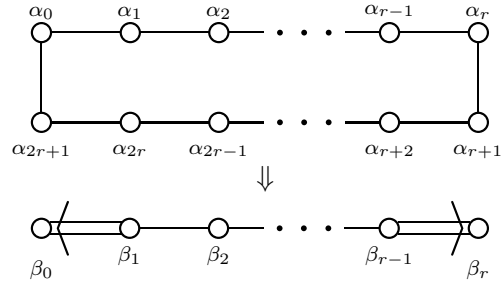

\end{example}

\begin{example}\label{exa:3}

Now choose $\mathfrak{g}\simeq \A_{2r}^{(1)} $ and fix up the

involution $\mathcal{C} $ by:
\begin{eqnarray}\label{eq:C_3}
&&\mathcal{C} (q_k) =-q_{2r+2-k}, \qquad \mathcal{C} (p_k)
=-p_{2r+2-k},\qquad k=1,\dots , r-1, \nonumber\\
&& \mathcal{C} (q_{r})=-q_{r}, \qquad \mathcal{C} (p_{r}) =-p_{r}.
\end{eqnarray}
The coordinates in $\mathcal{M}_\pm $:
\begin{eqnarray}\label{eq:p-pm3}
q_{k}^{\pm} = {1 \over \sqrt{2}} (q_k\mp q_{2r+2-k}),\qquad
p_{k}^{\pm} = {1 \over \sqrt{2}} (p_k\mp p_{2r+2-k}), k=1, \dots , r-1;\\
q_{r}^- = q_{r}, \qquad  p_{r}^- = p_{r},\nonumber
\end{eqnarray}
i.e., $\dim \mathcal{M}_+=2r-2 $ and  $\dim\mathcal{M}_-=2r $.

This involution induces an involution $\mathcal{C}^\# $ of the Kac-Moody
algebra $\A_{2r}^{(1)} $ which acts on the root space as follows
(see fig.~\ref{fig:3}):
\begin{eqnarray}\label{eq:E3.2}
\mathcal{C}^\# e_k =-e_{2r+2-k}, \qquad k=1,\dots, r; \qquad
k=r+2,\dots ,  2r+1; \\
\mathcal{C}^\#\alpha _k =\alpha _{2r+1-k}, \qquad k=1,\dots r; \qquad
\mathcal{C}^\# \alpha _0 =\alpha _0,
\end{eqnarray}
The involution $\mathcal{C}^\# $ splits the root space $\bbbe^{2r}
$ into a direct sum of its eigensubspaces:
$\bbbe^{2r}=\bbbe_+\oplus \bbbe_- $ with $\dim \bbbe_+=r-1$ and
$\dim \bbbe_-=r $. The restriction of $\pi $ onto $\bbbe_+ $ leads
to the admissible root system $\pi'=\{ \beta _0,\dots, \beta _r\}
$ of $\A_{2r}^{(2)} $:
\begin{equation}\label{eq:bet-3}
\beta _k={1 \over 2} (\alpha _k+\mathcal{C}^\#\alpha _k ), \qquad
k=1,\dots, r-1; \qquad \beta _0=\alpha _0, \qquad \beta _r=\alpha _r.
\end{equation}
The subspace $\bbbe_- $ is spanned by
\begin{equation}\label{eq:gam-3}
\gamma _k={1 \over 2} (\alpha _k-\mathcal{C}^\#\alpha _k ), \qquad
k=1,\dots, r.
\end{equation}

Then the densities $\mathcal{H}_{3}^\bbbr$,
$\omega _{3}^\bbbr$  for the RHF of AFTF equal:
\begin{eqnarray}\label{eq:H3}
&& \mathcal{H}_{3}^\bbbr  = {1\over 2} \sum_{k=1}^{r-1}
p_{k}^{+}{}^2 - {1\over 2} \sum_{k=1}^{r} p_{k}^{-}{}^2 + 2
e^{-q_{r-1}^+/\sqrt{2}} \cos \left(
{q_{r-1}^- \over \sqrt{2} } - q_{r+1}^-\right) \nonumber\\
&& +\sum_{k=1}^{r-1} 2 e^{ (q_{k+1}^+ -q_k^+)/\sqrt{2}} \cos
\left( {q_{k+1}^- -q_k^- \over \sqrt{2}}\right)+ 2
e^{q_1^+/\sqrt{2}} \cos \left( {q_{1}^- \over \sqrt{2} } -
q_{r+1}^-\right) .\nonumber \\
\label{eq:ome3} && \omega _{3}^\bbbr = \sum_{k=1}^{r-1} \delta
p_k^+\wedge \delta q_k^+ - \sum_{k=1}^{r} \delta p_k^-\wedge
\delta q_k^-
\end{eqnarray}
where $\pb_k^\pm = d\q_k^\pm /dx $. If we put $q_j^- = 0 $ and
$p_j^-=0 $ we get the reduced ATFT related to the Kac-Moody
algebra $\A_{2r}^{(2)} $ \cite{SasKha}.

\begin{figure}[htb]
\begin{picture}(300,120)
\put(111,83){$\lronit{\alp{0}}.$}
\put(142,96){$\ncddrdu{\alp{1}}.{\alp{2}}.{\alp{r-2}}.
     {\alp{r-1}}.{\alp{r}}.$}
\put(142,45){$\ncddrud{\alp{2r}}.{\alp{2r-1}}.{\alp{2r-2}}.
     {\alp{r+2}}.{\alp{r+1}}.$}
\put(222,27){$\Downarrow$}
\put(109,5){$\datwon{\bet{0}}.{\bet{1}}.{\bet{2}}.{\bet{r-2}}.
     {\bet{r-1}}.{\bet{r}}.$}
\end{picture}
\caption{$\A_{2r}^{(1)}\rightarrow \A_{2r}^{(2)}$}
\label{fig:3}
\end{figure}

\end{example}

\begin{example}\label{exa:4}

Let us choose now ${\frak g}\simeq {\bf D}_{r+1}^{(1)}$ and fix up the
involution $\mathcal{C} $ acting on the coordinates in the phase space by:
\begin{eqnarray}\label{eq:ex4}
\mathcal{C} (q_k) =q_{k}, \quad \mathcal{C} (p_k) =p_{k},\quad
k=1,\dots , r-1; \quad \mathcal{C} (q_{r})=-q_{r}, \quad
\mathcal{C} (p_{r}) =-p_{r}.
\end{eqnarray}
Then introducing on $\mathcal{M}_\pm$ new coordinates by
\begin{eqnarray}\label{eq:p-pm4}
&& q_k^+=q_k, \qquad  p_k^+=p_k, \qquad q_r^-=q_r, \qquad  p_r^-=p_r,
\qquad k=1,...,r;\\
&& q_{r+1}^{-} =q_{r+1}, \qquad p_{r+1}^{-} =p_{r+1},
\end{eqnarray}
i.e. $\dim \mathcal{M}_+=2r $ and  $\dim \mathcal{M}_-=2$.

This involution induces an involution $\mathcal{C}^\# $ of the Kac-Moody
algebra $\D_{r+1}^{(1)} $ which acts on the root space as follows
(see fig. \ref{fig:4}):
\begin{eqnarray}\label{eq:E4.2}
\mathcal{C}^\# e_k =e_{k}, \qquad k=1,\dots, r; \qquad
\mathcal{C}^\# e_{r+1} =-e_{r+1},\\
\mathcal{C}^\#\alpha _{k} =\alpha _{k}, \qquad
\mathcal{C}^\# \alpha _{r+1} =\alpha _r, \qquad
\mathcal{C}^\# \alpha _{r} =\alpha _{r+1},
\end{eqnarray}
The involution $\mathcal{C}^\# $ splits the root space
$\bbbe^{r+1} $ into a direct sum of its eigensubspaces:
$\bbbe^{r+1}=\bbbe_+\oplus \bbbe_- $ with $\dim \bbbe_+=r $ and
$\dim \bbbe_-=1 $. The restriction of $\pi $ onto $\bbbe_+ $ leads
to the admissible root system $\pi'=\{ \beta _0,\dots, \beta _r\}
$ of $\B_{r}^{(1)} $:
\begin{equation}\label{eq:bet-4}
\beta _k=\alpha _k, \qquad k=0,\dots , r-1; \qquad
\beta _r={1 \over 2} (\alpha _r+\mathcal{C}^\#\alpha _r ), \qquad
\end{equation}
The subspace $\bbbe_- $ is spanned by the only nontrivial vector
\begin{equation}\label{eq:gam-4}
\gamma _r={1 \over 2} (\alpha _r-\mathcal{C}^\#\alpha _r ) = -e_{r+1},
\end{equation}

The RHF is described by the following densities
$\mathcal{H}_{4}^\bbbr$, $\omega _{4}^\bbbr$:
\begin{eqnarray}\label{eq:H4}
&& \mathcal{H}_{4}^\bbbr  = {1\over 2} \sum_{k=1}^{r}
p_{k}^{+}{}^2 - {1\over 2}  p_{r+1}^{-}{}^2 +\sum_{k=0}^{r-1}
e^{q_{k+1}^+ -q_k^+} + 2 e^{q_{r-1}^+} \cos q_{r+1}^- ,\nonumber \\
&& \omega _{4}^\bbbr = \sum_{k=1}^{r} \delta p_k^+\wedge \delta
q_k^+ - \delta p_{r+1}^-\wedge \delta q_{r+1}^-.
\end{eqnarray}

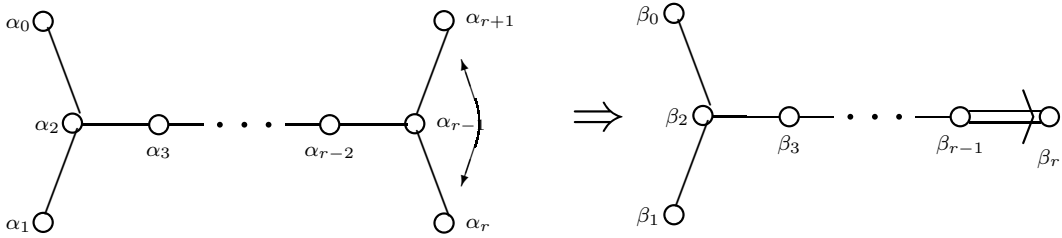
\begin{figure}[htb]
\begin{picture}(400,120)
\put(20,55){$\edddnd{\alp{0}}.{\alp{2}}.{\alp{1}}.{\alp{3}}.{\alp{r-2}}.
{\alp{r-1}}.{\alp{r}}.{\alp{r+1}}.~\Rightarrow~
\eddbnd{\bet{0}}.{\bet{2}}.{\bet{1}}.{\bet{3}}.{\bet{r-1}}.{\bet{r}}.$}
\put(192,67){\vector(-1,3){5}}
\put(192,49){\vector(-1,-3){5}}
\qbezier(192,67)(195,58)(192,49)
\end{picture}
\caption{Reductions of $\D_{r+1}^{(1)} $ to $\B_{r}^{(1)} $.}\label{fig:4}
\end{figure}

\end{example}

\begin{example}\label{exa:5}

Next we choose ${\frak g}\simeq {\bf D}_{r+2}^{(1)}$ and fix up the
involution $\mathcal{C} $ acting on the coordinates in the phase space by:
\begin{eqnarray}\label{eq:ex5}
\mathcal{C} (q_k) =q_{k}, \qquad \mathcal{C} (p_k) =p_{k},\qquad
k=2,\dots , r+1; \\
\mathcal{C} (q_{1})=-q_{1}, \qquad \mathcal{C} (p_{1}) =-p_{1}, \qquad
\mathcal{C} (q_{r+2})=-q_{r+2}, \qquad \mathcal{C} (p_{r+2}) =-p_{r+2}.
\end{eqnarray}
Then introducing on $\mathcal{M}_\pm$ new coordinates by
\begin{eqnarray}\label{eq:p-pm5}
&& q_k^+=q_k, \qquad  p_k^+=p_k, \qquad q_k^-=0, \qquad  p_k^-=0,
\qquad k=2,...,r+1;\\
&& q_1^-=q_1, \qquad  p_1^-=p_1, \qquad
q_{r+2}^{-} =q_{r+2}, \qquad p_{r+2}^{-} =p_{r+2},
\end{eqnarray}
i.e. $\dim \mathcal{M}_+=2r$ and  $\dim \mathcal{M}_-=4$.

This involution induces an involution $\mathcal{C}^\# $ of the Kac-Moody
algebra $\D_{r+2}^{(1)} $ which acts on the root space as follows
(see fig. \ref{fig:5}):
\begin{eqnarray}\label{eq:E5.2}
\mathcal{C}^\# e_k =e_{k}, \qquad k=2,\dots, r; \qquad
\mathcal{C}^\# e_{1} =-e_{1},\qquad \mathcal{C}^\# e_{r+1} =-e_{r+1},\\
\mathcal{C}^\#\alpha _{k} =\alpha _{k}, \qquad
\mathcal{C}^\# \alpha _{r+1} =\alpha _r, \qquad
\mathcal{C}^\# \alpha _{r} =\alpha _{r+1}, \nonumber\\
\mathcal{C}^\# \alpha _{1} =\alpha _0, \qquad
\mathcal{C}^\# \alpha _{0} =\alpha _{1},
\end{eqnarray}
The involution $\mathcal{C}^\# $ splits the root space
$\bbbe^{r+2} $ into a direct sum of its eigensubspaces:
$\bbbe^{r+2}=\bbbe_+\oplus \bbbe_- $ with $\dim \bbbe_+=r $ and
$\dim \bbbe_-=2 $. The restriction of $\pi $ onto $\bbbe_+ $ leads
to the admissible root system $\pi'=\{ \beta _0,\dots, \beta _r\}
$ of $\D_{r+1}^{(2)} $:
\begin{equation}\label{eq:bet-5}
\beta _k=\alpha _k, \qquad k=1,\dots , r-1; \qquad
\beta _j={1 \over 2} (\alpha _j+\mathcal{C}^\#\alpha _j ), \qquad j=0,r+1.
\end{equation}
The subspace $\bbbe_- $ is spanned by the vectors
\begin{equation}\label{eq:gam-5}
\gamma _{r+1}={1 \over 2} (\alpha _r-\mathcal{C}^\#\alpha _r ) = -e_{r+2},
\qquad  \gamma _{1}={1 \over 2} (\alpha _1-\mathcal{C}^\#\alpha _0 ) =
e_{1},
\end{equation}

The RHF is described by the following densities
$\mathcal{H}_{5}^\bbbr$, $\omega _{5}^\bbbr$:
\begin{eqnarray}\label{eq:H5}
&& \mathcal{H}_{5}^\bbbr  = {1\over 2} \sum_{k=2}^{r+1}
p_{k}^{+}{}^2 - {1\over 2} (p_1^-{}^2 + p_{r+2}^{-}{}^2) +\sum_{k=1}^{r}
2e^{-(\q_+, \beta _k)} + e^{-(\q_+, \beta _0)} \cos q_{1}^- \\
&& + e^{-(\q_+, \beta _{r})} \cos q_{r+2}^-,\nonumber \\
&& \omega _{5}^\bbbr = \sum_{k=2}^{r+1} \delta p_k^+\wedge \delta q_k^+ -
\delta p_1^-\wedge \delta q_1^- -  \delta p_{r+2}^-\wedge \delta
q_{r+2}^-.
\end{eqnarray}

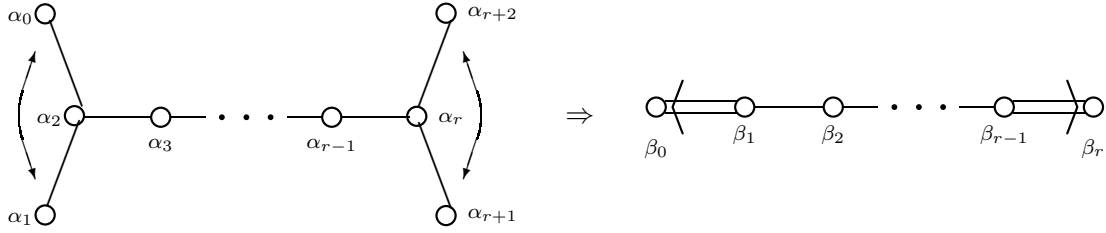
\begin{figure}
\begin{picture}(400,120)
\put(20,55){$\edddnd{\alp{0}}.{\alp{2}}.{\alp{1}}.{\alp{3}}.{\alp{r-1}}.
{\alp{r}}.{\alp{r+1}}.{\alp{r+2}}.~\Rightarrow~
\edddniid{\bet{0}}.{\bet{1}}.{\bet{2}}.{\bet{r-1}}.{\bet{r}}.$}
\put(192,67){\vector(-1,3){5}}
\put(192,49){\vector(-1,-3){5}}
\qbezier(192,67)(195,58)(192,49)
\put(20,67){\vector(1,3){5}}
\put(20,49){\vector(1,-3){5}}
\qbezier(20,67)(17,58)(20,49)
\end{picture}
\caption{Reductions of $\D_{r+2}^{(1)} $ to $\D_{r+1}^{(2)} $.}\label{fig:5}
\end{figure}

\end{example}

\begin{example}\label{exa:6}

Next we choose ${\frak g}\simeq {\bf D}_{2r}^{(1)}, r>3$ and fix
up the involution $\mathcal{C} $ acting on the coordinates in the
phase space by:
\begin{eqnarray}\label{eq:ex6}
\mathcal{C} (q_k) =-q_{2r+1-k}, \qquad \mathcal{C} (p_k) =
-p_{2r+1-k},\qquad k=1,\dots , r.
\end{eqnarray}
Then introducing on $\mathcal{M}_\pm$ new coordinates by
\begin{eqnarray}\label{eq:p-pm6}
&& q_k^+= {1 \over \sqrt{2} }(q_k -q_{2r+1-k}), \qquad
p_k^+= {1 \over \sqrt{2} }(p_k -p_{2r+1-k}), \qquad k=1,...,r;\\
&& q_k^-= {1 \over \sqrt{2} }(q_k +q_{2r+1-k}), \qquad
p_k^-= {1 \over \sqrt{2} }(p_k +p_{2r+1-k}), \qquad k=1,...,r;
\end{eqnarray}
i.e. $\dim \mathcal{M}_+=2r $ and  $\dim \mathcal{M}_-=2r$.

This involution induces an involution $\mathcal{C}^\# $ of the Kac-Moody
algebra $\D_{2r}^{(1)} $ which acts on the root space as follows
(see fig. \ref{fig:6}):
\begin{eqnarray}\label{eq:E6.2}
\mathcal{C}^\# e_k =-e_{2r+1-k}, \qquad k=1,\dots, 2r; \\
\mathcal{C}^\#\alpha _{k} =\alpha _{2r-k}, \qquad k=1,\dots, 2r.
\end{eqnarray}
The involution $\mathcal{C}^\# $ splits the root space $\bbbe^{2r}
$ into a direct sum of its eigensubspaces:
$\bbbe^{2r}=\bbbe_+\oplus \bbbe_- $ with $\dim \bbbe_+=r $ and
$\dim \bbbe_-=r $. The restriction of $\pi $ onto $\bbbe_+ $ leads
to the admissible root system $\pi'=\{ \beta _0,\dots, \beta _r\}
$ of $\A_{2r-1}^{(2)} $:
\begin{equation}\label{eq:bet-6}
\beta _k={1 \over 2} (\alpha _k+\mathcal{C}^\#\alpha _k ), \qquad
k=1,\dots , r-1; \qquad
\beta _r=\alpha _r.
\end{equation}
The subspace $\bbbe_- $ is spanned by the vectors
\begin{equation}\label{eq:gam-6}
\gamma _k={1 \over 2} (\alpha _k-\mathcal{C}^\#\alpha _k ), \qquad
k=1,\dots , r-1.
\end{equation}

The RHF is described by the following densities
$\mathcal{H}_{6}^\bbbr$, $\omega _{6}^\bbbr$:
\begin{eqnarray}\label{eq:H6}
&& \mathcal{H}_{6}^\bbbr  = {1\over 2} \sum_{k=2}^{r+1}
(p_{k}^{+}{}^2 -p_{k}^{+}{}^2)
+\sum_{k=1}^{r-1} 2e^{-(\q_+, \beta _k)}\cos (\q_-, \gamma _k) \\
&& + e^{-(\q_+, \beta _0)} \cos (\q_-, \gamma _0) + e^{-(\q_+, \beta _1)}
\cos (\q_-, \gamma _1) + e^{-(\q_+, \beta _r)} \cos (\q_-, \gamma _r),
\nonumber\\
&& \omega _{6}^\bbbr = \sum_{k=1}^{r}
(\delta p_k^+\wedge \delta q_k^+ - \delta p_k^-\wedge \delta q_k^-).
\end{eqnarray}

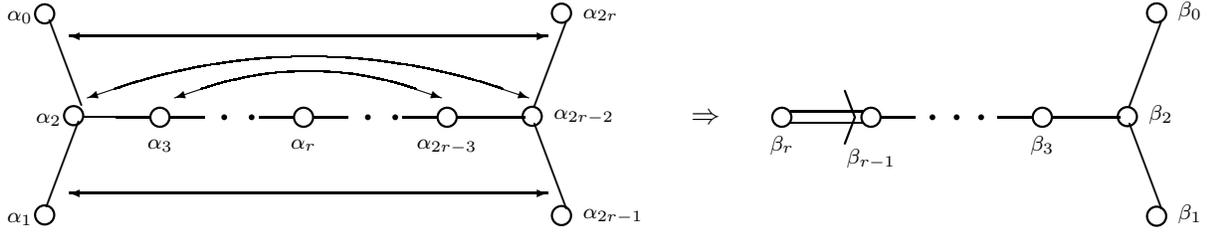
\begin{figure}
\begin{picture}(400,120)
\put(5,55){$\edddnds{\alp{0}}.{\alp{2}}.{\alp{1}}.{\alp{3}}.{\alp{r}}.
{\alp{2r-3}}.{\alp{2r-2}}.
{\alp{2r-1}}.{\alp{2r}}.~\Rightarrow~
\eddanod{\bet{r}}.{\bet{r-1}}.{\bet{3}}.{\bet{2}}.
           {\bet{1}}.{\bet{0}}.$}
\put(73,68){\vector(-3,-1){10}}
\put(154,68){\vector(3,-1){10}}
\qbezier(73,68)(113.5,81.5)(154,68)
\put(40,68){\vector(-3,-1){10}}
\put(187,68){\vector(3,-1){10}}
\qbezier(40,68)(113.5,92.5)(187,68)
\put(197,88){\vector(1,0){7}}
\put(30,88){\vector(-1,0){7}}
\put(30,88){\line(1,0){167}}
\put(30,28.5){\vector(-1,0){7}}
\put(197,28.5){\vector(1,0){7}}
\put(30,28.5){\line(1,0){167}}
\end{picture}
\caption{Reductions of $\D_{2r}^{(1)} $ to $\A_{2r-1}^{(2)}
$.}\label{fig:6} \end{figure}

\end{example}

\begin{example}\label{exa:7}

Finally we choose ${\frak g}\simeq {\bf D}_{2r+1}^{(1)}, r>3$ and
fix up the involution $\mathcal{C} $ acting on the coordinates in
the phase space by:
\begin{eqnarray}\label{eq:ex7}
\mathcal{C} (q_k) =-q_{2r+2-k}, \qquad \mathcal{C} (p_k) =
-p_{2r+2-k},\qquad k=1,\dots , r; \\
\mathcal{C} (q_{r+1})=-q_{r+1}, \qquad \mathcal{C} (p_{r+1}) =
-p_{r+1}.
\end{eqnarray}
Then introducing on $\mathcal{M}_\pm$ new coordinates by
\begin{eqnarray}\label{eq:p-pm7}
&& q_k^+= {1 \over \sqrt{2} }(q_k -q_{2r+2-k}), \qquad
p_k^+= {1 \over \sqrt{2} }(p_k -p_{2r+2-k}), \qquad k=1,...,r;\\
&& q_k^-= {1 \over \sqrt{2} }(q_k +q_{2r+2-k}), \qquad
p_k^-= {1 \over \sqrt{2} }(p_k +p_{2r+2-k}), \qquad k=1,...,r;\\
&& q^-_{r+1}=q_{r+1}, \qquad p^-_{r+1}=p_{r+1},
\end{eqnarray}
i.e. $\dim \mathcal{M}_+=2r $ and  $\dim \mathcal{M}_-=2r+2$.

This involution induces an involution $\mathcal{C}^\# $ of the Kac-Moody
algebra $\D_{2r+1}^{(1)} $ which acts on the root space as follows
(see fig. \ref{fig:7}):
\begin{eqnarray}\label{eq:E7.2}
\mathcal{C}^\# e_k =-e_{2r+2-k}, \qquad k=1,\dots, 2r+1; \\
\mathcal{C}^\#\alpha _{k} =\alpha _{2r+1-k}, \qquad k=0,\dots, 2r.
\end{eqnarray}
The involution $\mathcal{C}^\# $ splits the root space
$\bbbe^{2r+1} $ into a direct sum of its eigensubspaces:
$\bbbe^{2r+1}=\bbbe_+\oplus \bbbe_- $ with $\dim \bbbe_+=r $ and
$\dim \bbbe_-=r +1$. The restriction of $\pi $ onto $\bbbe_+ $
leads to the admissible root system $\pi'=\{ \beta _0,\dots, \beta
_r\} $ of $\B_{r}^{(1)} $:
\begin{equation}\label{eq:bet-7}
\beta _k={1 \over 2} (\alpha _k+\mathcal{C}^\#\alpha _k ), \qquad
k=0,\dots , r.
\end{equation}
The subspace $\bbbe_- $ is spanned by the vectors $ \gamma _k={1
\over 2} (\alpha _k-\mathcal{C}^\#\alpha _k ), \, k=1,\dots , r+1.
$ The RHF is described by the following densities
$\mathcal{H}_{7}^\bbbr$, $\omega _{7}^\bbbr$:
\begin{eqnarray}\label{eq:H7}
&& \mathcal{H}_{7}^\bbbr  = {1\over 2} \sum_{k=1}^{r}
p_{k}^{+}{}^2 - {1\over 2} \sum_{k=1}^{r+1}p_{k}^{-}{}^2
+\sum_{k=2}^{r} 4e^{-(\q_+, \beta _k)}\cos (\q_-, \gamma _k) \\
&&  +2e^{-(\q_+, \beta _0)} \cos (\q_-, \gamma _0) +
2e^{-(\q_+, \beta _1)} \cos (\q_-, \gamma _1) ,\nonumber\\
&& \omega _{7}^\bbbr = \sum_{k=1}^{r}
\delta p_k^+\wedge \delta q_k^+ - \sum_{k=1}^{r+1} \delta p_k^-\wedge
\delta q_k^-.
\end{eqnarray}

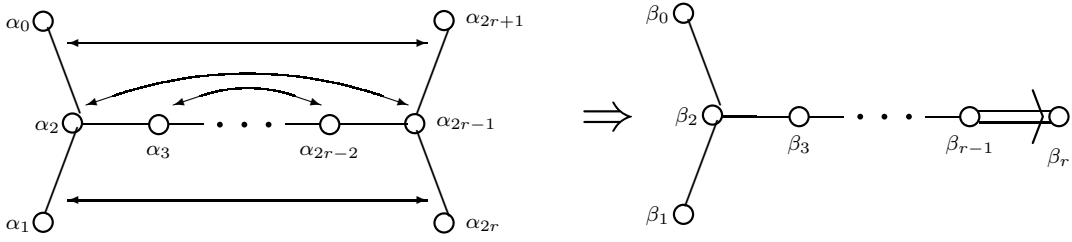
\begin{figure}
\begin{picture}(400,120)
\put(20,50){$\edddnd{\alp{0}}.{\alp{2}}.{\alp{1}}.{\alp{3}}.{\alp{2r-2}}.
{\alp{2r-1}}.{\alp{2r}}.{\alp{2r+1}}.~\Rightarrow~
\eddbnd{\bet{0}}.{\bet{2}}.{\bet{1}}.{\bet{3}}.{\bet{r-1}}.{\bet{r}}.$}
\put(88,63){\vector(-3,-1){10}}
\put(124,63){\vector(3,-1){10}}
\qbezier(88,63)(106,69)(124,63)
\put(55,63){\vector(-3,-1){10}}
\put(157,63){\vector(3,-1){10}}
\qbezier(55,63)(106,80)(157,63)
\put(167,83){\vector(1,0){7}}
\put(45,83){\vector(-1,0){7}}
\put(45,83){\line(1,0){122}}
\put(45,23.5){\vector(-1,0){7}}
\put(167,23.5){\vector(1,0){7}}
\put(45,23.5){\line(1,0){122}}
\end{picture}
\caption{Reductions of $\D_{2r+1}^{(1)} $ to $\B_{r}^{(1)}
$.}\label{fig:7}
\end{figure}

\end{example}

\section{Conclusions}

We outlined the generalization of the notion of RHF to infinite
dimensional Hamiltonian systems using as a basis the ATFT type models. We
established that the special properties of these models allow us to relate
the construction of the RHF to the study of $\bbbz_2 $ symmetries of the
extended Dynkin diagrams. The general construction is illustrated by
several examples of such models related to the type ${\bf A}_n $ and ${\bf
D}_n$ Kac-Moody algebras.

The ATFT models obtained here are generalizations of the ATFT in
\cite{SasKha}. Indeed, they contain two types of fields $\q_+(x,t) $ and
$\q_-(x,t) $ which have different properties with respect to the
involution $\mathcal{C} $, namely $\mathcal{C}\q_\pm(x,t) =\pm q_\pm(x,t)
$. The $\bbbz_2 $-reductions analyzed in \cite{SasKha} contain only one
type of fields invariant with respect to $\mathcal{C} $.

Some additional problems are natural extensions to the results presented
here. The first one is the complete classification of all RHF of ATFT,
including the cases related to the exceptional Kac-Moody algebras. Next is
the derivation of their Hamiltonian properties based on the classical $R
$-matrix approach, see \cite{PPK}. The last and technically more involved
problem is to solve the inverse scattering problem for the Lax operator
and thus prove the complete integrability of all these models.

In \cite{GIG} it was shown that the dynamics of different RHF for
the finite-dimensional Toda systems may be qualitatively
different; for example some of the non-compact trajectories may
become compact. Similar qualitative changes may be expected also
in the infinite-dimensional cases.

Another interesting problem is related to the fact that the
integrable systems possess a hierarchy of Hamiltonian structures,
for review of the infinite-dimensional cases see e.g.
\cite{DrSok,Holy} and the references therein;  for the
(finite-dimensional) Toda chains see \cite{Dam1,Dam2}. It is an
open problem to construct the RHF for ATFT using some of its
higher Hamiltonian structures.

Finally, this method provides a tool for systematic
construction and classification of the RHF of a wide class of finite and
infinite-dimensional Hamiltonian systems, which need not be necessarily
integrable.  Therefore to each Hamiltonian system one can associate a
family of its RHF. To do this one has to construct the corresponding
involutions of the Poisson brackets preserving the Hamiltonian.

\begin{acknowledgments}
One of us (GGG) thanks the organizing committee of the European
Research Conference on ``Symmetries and Integrability of
Difference Equations'' (SIDE VI) for the scholarship and for the
warm hospitality in Helsinki and  the  Expert Council for Young
Scientists of the Bulgarian Academy of Sciences for financial
support.  We also acknowledge support by the National Science
Foundation of Bulgaria, contract No. F-1410.
\end{acknowledgments}

\label{lastpage}

\end{document}